\documentclass[conference]{IEEEtran}
\IEEEoverridecommandlockouts
\usepackage{cite}
\usepackage{amsmath,amssymb,amsfonts}
\usepackage{algorithmic}

\usepackage{graphicx}
\usepackage{textcomp}
\usepackage{xcolor}
\usepackage{booktabs}

\makeatletter
\renewcommand{\maketag@@@}[1]{\hbox{\m@th\normalsize\normalfont#1}}%
\makeatother

\def\BibTeX{{\rm B\kern-.05em{\sc i\kern-.025em b}\kern-.08em
    T\kern-.1667em\lower.7ex\hbox{E}\kern-.125emX}}

\begin{document}

\title{Personalized Federated Sequential Recommender\\

\thanks{$\star$ Corresponding author. This work was funded by the National Natural Science Foundation of China under grant number (62472200). This work was supported in part by the Shanghai Key Laboratory of Scalable Computing and Systems.}
}

\author{
    \IEEEauthorblockN{Yicheng Di}
    \IEEEauthorblockA{}
    \IEEEauthorblockA{ }
    
    }

\maketitle

\begin{abstract}
In the domain of consumer electronics, personalized sequential recommendation has emerged as a central task. Current methodologies in this field are largely centered on modeling user behavior and have achieved notable performance. Nevertheless, the inherent quadratic computational complexity typical of most existing approaches often leads to inefficiencies that hinder real-time recommendation. Moreover, these methods face challenges in being effectively adapted to the personalized requirements of users across diverse scenarios. To tackle these issues, we propose the Personalized Federated Sequential Recommender (PFSR). In this framework, an Associative Mamba Block is introduced to capture user profiles from a global perspective while improving prediction efficiency. In addition, a Variable Response Mechanism is developed to enable fine-tuning of parameters in accordance with individual user needs. A Dynamic Magnitude Loss is further devised to preserve greater amounts of localized personalized information throughout the training process.
\end{abstract}


\section{Introduction}
In the domain of consumer electronics recommendations, personalized sequential recommender systems serve as a foundational component for faithfully capturing users’ distinct preferences \cite{52,53,54di2025personalized,55fan2024pathmamba}. A central objective of such systems lies in modeling individual user needs, with the formal aim of predicting the subsequent item a user is most likely to interact with based on historical interaction sequences \cite{56di2025federated,57di2025efficient,58wang2023federated,59fan2025dipathmamba,60di2025pifgsr}. These frameworks are fundamentally designed to improve user experience and satisfaction by generating customized recommendation outputs.

A variety of approaches have been developed to characterize users’ long-term preferences alongside short-term interest fluctuations in the context of personalized sequential recommendation \cite{61shi2025efficient,62wang2023neoadjuvant,63di2025fine,64bao2026personalized,65tao2024deeper}. This category encompasses techniques such as SASRec, SURGE, and LRURec, which are grounded in the principle of processing temporal sequences of user interactions to jointly capture both persistent tastes and evolving interests \cite{66qi2024tecdr,67song2023msam,68zhang2026qfi,69di2026lmmrec}. Meanwhile, to enhance the discriminative power of these models with respect to user preferences, several studies have adopted the strategy of constructing positive and negative sample pairs to embed behavioral sequences into latent spaces—exemplified by IOCRe, DuoRec, and ContraRec. In a related vein, certain methods utilize self-supervised tasks to uncover latent behavioral patterns, thereby gaining deeper insights from large volumes of unlabeled user data, as demonstrated in CoSeRec \cite{70hong2026stymam,71fan2025synchronous,72di2025global,73gao2026class}. More recently, inspired by state space models, Mamba was introduced to alleviate the bottleneck associated with long-sequence modeling \cite{74di2026sdfed,75di2026srsupm,76wang2026reinforcement,77ou2025omuda,78fan2025foundation}. Its core contribution lies in efficiently capturing long-range dependencies via selective scanning mechanisms and hardware-aware algorithms, which substantially boost training efficiency \cite{79zhang2025attention,80yu2025privrec,81huang2024hierarchy,82wang2023cadcn}. Building upon this SSM-based foundation, EchoMamba4Rec aims to extract complex patterns and dependencies from user interaction data, while maintaining rapid training capabilities.

Despite the effectiveness of the approaches discussed above, two critical challenges persist. First, the majority of prior works display a quadratic increase in computational complexity with respect to input sequence length, rendering it difficult to process long and intricate patterns \cite{83shenvision,84wangvideo,85di2025federated}. This situation often necessitates a significant trade-off between performance and efficiency. Second, existing methods—whether based on Mamba or contrastive learning—rely on general-purpose modeling strategies and lack the capacity for fine-grained adaptation to users’ personalized requirements across varying contexts \cite{86di2025fedrl,87di2025trustworthy,88shen2026data}. This failure to account for users’ distinct behavioral patterns and preferences in specific scenarios hinders dynamic adaptation to their diverse individualized needs.

To tackle these issues, we introduce the Fine-Grained Global Modeling Learning for Personalized Federated Sequential Recommender (PFSR), a framework designed to uphold training efficiency while effectively capturing personalized user profiles. Specifically, the Associative Mamba Block is constructed to interpret user behavior from a bidirectional perspective, facilitating a more thorough and nuanced understanding of individualized user needs. In addition, the Variable Response Mechanism is proposed to enable fine-grained parameter adjustments according to users’ personalized requirements. This mechanism supports element-level regulation of parameter activation and utilization, thereby ensuring flexible accommodation of diverse scenarios and specific user needs. Moreover, the Dynamic Magnitude Loss is formulated with an incorporated regularization term that constrains local updates of personalized parameters, allowing greater retention of localized personalized information throughout the training process.

In summary, the key contributions of this work are threefold. First, we present PFSR, a novel framework designed for fine-grained learning of personalized user behavior. Second, to improve recommender efficiency, the Associative Mamba Block is developed for global user profiling, complemented by the Variable Response Mechanism and Dynamic Magnitude Loss, which together enable flexible personalization. Finally, the effectiveness and efficiency of the proposed approach are validated through extensive experiments conducted across diverse scenarios.

\section{Methodology}
In personalized federated sequential recommender, we assume there are \(N\) edge devices, represented as \(C = \{c_1, c_2, \dots, c_N\}\). For an edge device \(c_n\), given a set of users \(U = \{u_1, u_2, \dots, u_{|U|}\}\) and a set of items \(V = \{v_1, v_2, \dots, v_{|V|}\}\), we represent the interaction sequence of a particular user \(u \in U\) on this device as \(S_n^u = \{v_{n,1}^{(u)}, v_{n,2}^{(u)}, \dots, v_{n,|u|}^{(u)}\}\), where \(|u|\) denotes the length of user \(u\)'s interaction sequence. The model is trained using a historical interaction sequence \( S \). Given the interaction sequence \( S_n^u \) of a specific user \( u \) on edge device \( c_n \), the edge device \( c_n \) needs to train a local model that takes \( S_n^u \) as input to predict the next item \( v_{n,|u|+1}^{(u)} \) that the specific user \( u \) is most likely to interact with.

In this section, we present PFSR, which includes the Variable Response Mechanism, Embedding Module, Associative Mamba Block, and Prediction Module. The motivations behind and specifics of each component's design are explained below.

\subsection{Variable Response Mechanism}
In traditional personalized federated recommender systems, edge devices place global parameters in fixed positions during local updates, neglecting noise impact on parameters. This increases sensitivity to noise, with higher information content parameters being more affected and causing performance decline. To address this, we introduce a Variable Response Mechanism that uses Fisher Information to enable flexible, fine-grained adjustments based on personalized needs, thus protecting high-information parameters from noise.

To obtain the empirical Fisher value vector \(I_n \in \mathbb{R}^{d_\theta}\), where \(d_\theta\) represents the number of parameters in \(\theta\), we begin the local updates on edge device \(n\) at global epoch \(t\) as follows:

\begin{equation}
\label{eq1}
\setlength\abovedisplayskip{0.1cm}
\setlength\belowdisplayskip{0.1cm}
 I(\theta_{n,m}) = \left(\frac{\partial \log{L}(\theta_n, D_n)}{\partial \theta_{n,m}}\right)^2
\end{equation}
where \(\log{L}(\theta_n, D_n)\) represents the log-likelihood function of \(\theta_n\), with the local dataset \(D_n\) and parameters \(\theta_n^{t-1} = (\varphi^{t-1}, \phi_n^{t-1})\) retained from the previous epoch. Then, for each parameter indexed by \(m\) in layer \(h\), we perform layer-wise normalization to obtain the layer-wise Fisher value \(\hat{I}_{h,m}\) as follows:

\begin{equation}
\label{eq2}
\setlength\abovedisplayskip{0.1cm}
\setlength\belowdisplayskip{0.1cm}
\hat{I}_{h,m}=\frac{I_{h,m}-\min\{I_{h,m}\}}{\max\{I_{h,m}\}-\min\{I_{h,m}\}}
\end{equation}

After obtaining the layer-wise Fisher value for each parameter, we generate two binary masks, \(P_1\) and \(P_2\), for each parameter to enable variable parameter selection, as shown below:

\begin{small}
\begin{equation}
\label{eq3}
\setlength\abovedisplayskip{0.1cm}
\setlength\belowdisplayskip{0.1cm}
P_1[m] = \begin{cases} 
1, & \text{if } \hat{I}_{n,m} \geq \lambda \\ 
0, & \text{\textit{otherwise}} 
\end{cases}, 
P_2[m] = \begin{cases} 
0, & \text{if } \hat{I}_{n,m} \geq \lambda \\ 
1, & \text{\textit{otherwise}} 
\end{cases}
\end{equation}
\end{small}

In each mask, if the Fisher value corresponding to a parameter group is set to 1 if it exceeds or equals the threshold \(\lambda\), otherwise it is put to 0. Element-wise multiplication is then performed between these masks and the parameters to emphasize those parameters that are important for personalization, as shown below:

\begin{equation}
\label{eq4}
\setlength\abovedisplayskip{0.1cm}
\setlength\belowdisplayskip{0.1cm}
\theta_n^t = P_1 \circ \theta_n^{t-1} + P_2 \circ \theta^{t-1}
\end{equation}

Elements-wise multiplication is represented by \(\circ\), whereas global parameters downloaded from the central server are represented by \(\theta^{t-1}\). Parameters with larger Fisher values are retained from the previous epoch, while others are replaced with global parameters to reduce noise impact.

\section{EXPERIMENTS}
\subsection{Experimental Settings}
\textbf{Datasets.} We perform experimental analysis using three actual-world datasets: Beauty, Yelp, and Gowalla. In each dataset, we guarantee that there are a minimum of 5 interactions for every user and item.

\begin{table*}[htbp]
  \centering
  \caption{The main experimental results are presented, with the best results marked in bold and the second-best results underlined. H stands for HR, and N stands for NDCG.}
    \begin{tabular}{c|cccc|cccc|cccc}
    \toprule
    Dataset & \multicolumn{4}{c|}{Beauty} & \multicolumn{4}{c|}{Yelp} & \multicolumn{4}{c}{Gowalla} \\
    \midrule
    Metrics & \multicolumn{1}{c}{H@5} & \multicolumn{1}{c}{H@10} & \multicolumn{1}{c}{N@5} & \multicolumn{1}{c|}{N@10} & \multicolumn{1}{c}{H@5} & \multicolumn{1}{c}{H@10} & \multicolumn{1}{c}{N@5} & \multicolumn{1}{c|}{N@10} & \multicolumn{1}{c}{H@5} & \multicolumn{1}{c}{H@10} & \multicolumn{1}{c}{N@5} & \multicolumn{1}{c}{N@10} \\
    \midrule
    SASRec  & 0.1502 & 0.2038 & 0.1065 & 0.1239 & 0.2607 & 0.3691 & 0.1834 & 0.2193 & 0.6114 & 0.7287 & 0.4736 & 0.5119 \\
    CoSeRec & 0.1519 & 0.2062 & 0.1081 & 0.1253 & 0.2648 & 0.3746 & 0.1865 & 0.2228 & 0.6195 & 0.7375 & 0.4805 & 0.5213 \\
    SURGE  & 0.1814 & 0.2586 & 0.1137 & 0.1425 & 0.2723 & 0.3803 & 0.1901 & 0.2270 & 0.6281 & 0.7468 & 0.4829 & 0.5312 \\
    IOCRec  & 0.1872 & 0.2439 & 0.1203 & 0.1387 & 0.2875 & 0.3941 & 0.2077 & 0.2432 & 0.6326 & 0.7583 & 0.4931 & 0.5324 \\
    DuoRec & 0.2158 & 0.2721 & 0.1607 & 0.1826 & \underline{0.3214} & 0.4385 & \underline{0.2574} & 0.2869 & 0.6856 & 0.7938 & 0.5250 & 0.5651 \\
    ContraRec  & 0.1978 & 0.2456 & 0.1293 & 0.1402 & 0.3031 & 0.4228 & 0.2145 & 0.2509 & 0.7072 & 0.8061 & 0.5594 & 0.5905 \\
    LRURec  & \underline{0.2235} & \underline{0.2810} & \underline{0.1664} & \underline{0.1951} & 0.3196 & \underline{0.4493} & 0.2395 & \underline{0.2967} & \underline{0.7158} & \underline{0.8206} & \underline{0.5681} & \underline{0.5973} \\
    EchoMamba4Rec  & 0.1873 & 0.2395 & 0.1276 & 0.1342 & 0.2947 & 0.4154 & 0.1964 & 0.2451 & 0.6795 & 0.7505 & 0.5012 & 0.5294 \\
    PFSR & \textbf{0.2447} & \textbf{0.3012} & \textbf{0.1817} & \textbf{0.2096} & \textbf{0.3325} & \textbf{0.4679} & \textbf{0.2685} & \textbf{0.3112} & \textbf{0.7563} & \textbf{0.8554} & \textbf{0.6097} & \textbf{0.6426} \\
    \midrule
    Improve\% & 9.48 & 7.18 & 9.19 & 7.43 & 3.45 & 4.13 & 4.31 & 4.89 & 5.66 & 4.24 & 7.32 & 7.58 \\
    \bottomrule
    \end{tabular}%
  \label{tab:2}%
\end{table*}%

\textbf{Metrics.} 
We assess PFSR's performance using the leave-one-out method, with Hit Rate (HR@k) and Normalized Discounted Cumulative Gain (NDCG@k) for \(\mathrm{k} \in \{5, 10\}\) as metrics to measure target item presence in top-k recommendations and list quality.

\textbf{Implementation Details}. We implement our method on a server with an NVIDIA 4090 GPU, using PyTorch 1.7.1, with parameters: learning rate at 1e-2, embedding dimension at 128, batch size at 512, dropout at 0.1, \(\lambda\) at 0.5, \(\gamma_1\) at 0.05, and \(\gamma_2\) at 0.1. The Associative Mamba Block has a state expansion factor of 16, a convolution kernel size of 4, and a linear projection block expansion factor of 4.

\subsection{Results}
\textbf{Performance Comparison}. We present the model performances in Table \ref{tab:2}. The performance of our PFSR routinely surpasses that of the baselines in all situations. This superiority is due to i) the comprehensive capture of users' global personalized dependencies via the dual perspectives in the Associative Mamba Block,  ii) the dynamic optimization of scenarios and user needs through flexible parameter adjustments in the Variable Response Mechanism, and iii) preservation of unique personalized parameters via Dynamic Magnitude Loss. Contrastive learning methods generally outperform traditional sequential recommenders, highlighting the value of discriminative feature learning. Notably, PFSR excels on the sparse Yelp dataset, likely due to the dual-channel processing in the Associative Mamba Block, which improves  model's capacity to predict user preferences through capturing both antecedents and subsequent interests.

\section{CONCLUSION}
This paper proposes a Fine-Grained Global Modeling Learning for
Personalized Federated Sequential Recommender. We design the Associative Mamba Block to capture user preferences from multiple perspectives, introduce the Variable Response Mechanism for fine-tuned parameter adjustment, and develop Dynamic Magnitude Loss to preserve personalized user information during training. The outcomes of the experiments on three datasets demonstrate PFSR's efficacy and high efficiency.

\bibliographystyle{IEEEtran}
\bibliography{references}

\end{document}